\documentstyle{article}

\begin{document}

\title{\flushright{\small KL-TH / 02-09} \bigskip \bigskip \bigskip \bigskip
\bigskip \bigskip \\
\center{ A Map between $q-$deformed  and ordinary Gauge
Theories}\thanks{%
Supported by DAAD}}
\author{L. Mesref\thanks{%
Email: lmesref@physik.uni-kl.de}}
\date{Department of Physics, Theoretical Physics\\
University of Kaiserslautern, Postfach 3049\\
67653 Kaiserslautern, Germany}
\maketitle

\begin{abstract}
In complete analogy with Seiberg-Witten map defined in noncommutative
geometry we introduce a new map between a $q$-deformed gauge theory and an
ordinary gauge theory. The construction of this map is elaborated in order
to fit the Hopf algebra structure.

\newpage
\end{abstract}

\section{Introduction}

The concept of space-time as a space of commuting coordinates is perhaps too naive and must possibly be modified at the Planck scale. In their seminal paper Connes, Douglas and Schwarz \cite{douglas} have found that noncommutative geometry \cite{connes} arises naturally in string theory. Quantum groups \cite{drinfeld} provide another consistent mathematical framework to formulate physical theories on noncommutative spaces. They appeared first in the study of integrable systems \cite{faddeev} and are now applied to many branches of mathematics and physics. Although a tremendous amount of literature has been devoted to the study of quantum groups, we are still lacking a map which relates the quantum deformed gauge theories and the ordinary ones. In the present paper, we introduce such a map. This map is a quantum analog of the Seiberg-Witten map \cite{seiberg}. 

Let us first recall that Seiberg-Witten map was discovered in the contex of string theory where it emerged from 2D-$ \sigma $-model regularized in different ways. Seiberg and Witten have schown that the noncommutativity depends on the choice of the
regularization procedure: it appears in point-spliting regularization
whereas it is not present in the Pauli Villars regularization. This
observation led them to argue that there exist a map connecting the
noncommutative gauge fields and gauge transformation parameter to the
ordinary gauge field and gauge parameter. This map can be interpreted as an
expansion of the noncommutative gauge field in $\theta $. \ Along similar
lines, we introduce a new map between the $q$-deformed and undeformed gauge
theories. This map can be seen as an infinitesimal shift in the parameter $q$%
, and thus as an expansion of the deformed gauge field in $q=e^{i\,\eta
}=1+i\eta +o \left( \eta ^{2}\right) $.

This paper is organized as follows. In Sec. 2, we recall the Seiberg-Witten
map. In Sec. 3, we present the $SU_{q}\left( 2\right) $\ quantum group
techniques. In Sec. 4, we recall the Woronowicz differential calculus using
the adjoint representation $M_{b}^{\,\,\,\,a}$ of the group $SU_{q}\left(
2\right) $. In Sec. 5, we consider an hybrid structure consisting of a
noncommutative base space defined by a Moyal product and a $q$-deformed \
nonabelian gauge group. This structure allows us to define a map which
relate $\ $the $q$-deformed noncommutative and the ordinary gauge fields. We
close this paper by constructing a new map relating the $q$-deformed and
ordinary gauge fields. 

\section{The Seiberg-Witten map}

Let us recall that in the noncommutative geometry the space-time coordinates
$x^{i}$ are replaced by the Hermitian generators $\widehat{x}^{i}$ of a
noncommutative $C^{\ast }$-algebra of functions on space-time which obey the
following algebra

\bigskip

\begin{equation}
\left[ \widehat{x}^{i},\widehat{x}^{j}\right] =i\theta ^{i\,j},  
\end{equation}

\bigskip

where $\theta ^{i\,j}$ is real and antisymmetric with dimension of length
squared. In this context the ordinary product of two functions is replaced
by the Groenewold-Moyal star-product \cite{moyal}

\bigskip

\begin{eqnarray}
f\left( x\right) \star g\left( x\right) &=&f\left( x\right) \,\exp \left(
\frac{i}{2}\overleftarrow{\partial }_{i}\,\theta ^{i\,j}\overrightarrow{%
\partial }_{j}\right) \,g\left( x\right)  \nonumber \\
&=&f\left( x\right) \,g\left( x\right) +\frac{i}{2}\theta ^{i\,j}\,\partial
_{i}f\left( x\right) \,\partial _{j}g\left( x\right) +o \left(
\theta ^{2}\right) .  
\end{eqnarray}

\bigskip

For the ordinary Yang-Mills theory the infinitesimal gauge transformations and field
strength are given by

\bigskip

\begin{eqnarray}
\delta _{\lambda }\,A_{i} &=&\partial _{i}\lambda +i\left[ \lambda ,A_{i}%
\right]  \nonumber \\
F_{i\,j} &=&\partial _{i}A_{j}-\partial _{j}A_{i}-i\left[ A_{i},A_{j}\right]
\nonumber \\
\delta _{\lambda }\,F_{i\,j} &=&i\left[ \lambda ,F_{i\,j}\right] \,
\end{eqnarray}

\bigskip

where the symbols $ \left[ .,. \right] $ mean commutators.

\bigskip

For the noncommutative gauge theory we just replace the matrix
multiplication by the $\star $ product. The infinitesimal gauge transformations are given by

\bigskip

\begin{eqnarray}
\widehat{\delta }_{\widehat{\lambda }}\widehat{A}_{i} &=&\partial _{i}%
\widehat{\lambda }+i\,\widehat{\lambda }\star \widehat{A}_{i}-i\,\widehat{A}%
_{i}\star \widehat{\lambda }  \nonumber \\
\widehat{\delta }_{\widehat{\lambda }}\widehat{F}_{i\,j} &=&i\widehat{%
\lambda }\star \widehat{F}_{i\,j}-i\widehat{F}_{i\,j}\star \widehat{\lambda }
\end{eqnarray}

where

\bigskip

\begin{equation}
\widehat{F}_{i\,j}=\partial _{i}\widehat{A}_{j}-\partial _{j}\widehat{A}%
_{i}-i\,\widehat{A}_{i}\star \widehat{A}_{j}+i\,\widehat{A}_{j}\star 
\widehat{A}_{i}\,. 
\end{equation}

\bigskip

Using first order expansion in $\theta $ these relations give

\bigskip

\begin{eqnarray}
\widehat{\delta }_{\widehat{\lambda }}\,\widehat{A}_{i} &=&\partial _{i}%
\widehat{\lambda }+\widehat{\lambda }\,\widehat{A}_{i}-\widehat{A}_{i}\,%
\widehat{\lambda }-\frac{1}{2}\theta ^{jk}\partial _{j}\widehat{\lambda }%
\,\partial _{k}\widehat{A}_{i}+\frac{1}{2}\theta ^{jk}\partial _{k}\widehat{A%
}_{i}\partial _{j}\widehat{\lambda }\,+o\left( \theta ^{2}\right)   \nonumber
\\
\widehat{\delta }_{\widehat{\lambda }}\,\widehat{F}_{ij} &=&i\widehat{F}%
_{ij}\,\widehat{\lambda }-i\widehat{\lambda }\,\widehat{F}_{ij}\,-\frac{1}{2}%
\theta ^{kl}\partial _{k}\widehat{\lambda }\,\partial _{l}\widehat{F}_{ij}+%
\frac{1}{2}\theta ^{kl}\partial _{k}\widehat{F}_{ij}\partial _{l}\widehat{%
\lambda }+o\left( \theta ^{2}\right) . 
\end{eqnarray}

\bigskip

To ensure that an ordinary gauge transformation of $A$ by $\lambda $ is
equivalent to a noncommutative gauge transformation of $\ \widehat{A}$ by $%
\widehat{\lambda }$ \ Seiberg and Witten have proposed the following relation

\bigskip

\begin{equation}
\widehat{A}\left( A\right) +\widehat{\delta }_{\widehat{\lambda }}\,\widehat{%
A}\left( A\right) =\widehat{A}\left( A+\delta _{\lambda }A\right) . 
\end{equation}

\bigskip

They first worked the first order in $\theta $ and wrote

\bigskip

\begin{eqnarray}
\widehat{A} &=&A+A^{\prime }\left( A\right)  \nonumber \\
\widehat{\lambda }\left( \lambda ,A\right) &=&\lambda +\lambda ^{\prime
}\left( \lambda ,A\right) .  
\end{eqnarray}

\bigskip

Expanding (7) in powers of $\theta $ they found

\bigskip

\begin{equation}
A_{i}^{\prime }\,\left( A+\delta _{\lambda }A\right) -A_{i}^{\prime
}\,\left( A\right) -\partial _{i}\lambda ^{\prime }-i\left[ \lambda ^{\prime
},A_{i}\right] -i\left[ \lambda ,A_{i}^{\prime }\right] =-\frac{1}{2}\theta
^{kl}\left( \partial _{k}\lambda \partial _{l}A_{i}+\partial
_{l}A_{i}\partial _{k}\lambda \right) +o \left( \theta ^{2}\right) ,
\end{equation}

\bigskip

where they used the expansion

\bigskip

\begin{equation}
f\star g=f\,g+\frac{1}{2}i\,\theta ^{i\,j}\partial _{i}f\,\partial _{j}g+%
o \left( \theta ^{2}\right) . 
\end{equation}

\bigskip

The equation (9) is solved by

\bigskip

\begin{eqnarray}
\widehat{A}_{i}\left( A\right) &=&A_{i}+A_{i}^{\prime }\left( A\right)
=A_{i}-\frac{1}{4}\theta ^{k\,l}\left\{ A_{k},\partial
_{l}A_{i}+F_{li}\right\} +o \left( \theta ^{2}\right)  \nonumber \\
\widehat{\lambda }\left( \lambda ,A\right) &=&\lambda +\lambda ^{\prime
}\left( \lambda ,A\right) =\lambda +\frac{1}{4}\theta ^{i\,j}\left\{
\partial _{i}\lambda ,A_{j}\right\} +o \left( \theta ^{2}\right) 
\end{eqnarray}

\bigskip

The equations (11) are called the Seiberg Witten map.

\bigskip

\section{The quantum group $SU_{q}\left( 2\right) $}

Let $\mathcal{A}$ be the associative \ unital $C-$algebra generated by the
linear transformations $M^{n}\,_{m}$ $\left( n,m=1,2\right) $

\bigskip

\begin{equation}
M^{n}\,_{m}=\left( 
\begin{array}{cc}
a & b \\ 
c & d
\end{array}
\right) , 
\end{equation}

\bigskip

the elements $a$, $b$, $c$, $d$ satisfying the relations

\bigskip

\begin{eqnarray}
ab &=&qba\quad \quad bc=cb\quad \quad \quad \quad \quad ac=qca,  \nonumber \\
bd &=&qdb\quad cd=qdc\quad ad-da=\left( q-q^{-1}\right) bc  
\end{eqnarray}

\bigskip

where $q$ is a deformation parameter. The classical case is obtained by
setting $q$ equal to one.

The $U_{q}\left( 2\right) $ is obtained by requiring that the unitary
condition hold for this $2\times 2$ \ quantum matrix:

\bigskip

\begin{equation}
M_{m}^{n\ \dagger }=M_{m}^{n\ -1}. 
\end{equation}

\bigskip

The $2\times 2$ matrix belonging to $U_{q}\left( 2\right) $ preserves the \
nondegenerate bilinear form \cite{dubois} $B_{nm}$

\bigskip

\begin{equation}
B_{nm}M_{k}^{n}M_{l}^{m}=D_{q}B_{kl},\quad B^{nm}M_{n}^{k}M_{m}^{l}=D_{q}B^{kl},\quad
B_{kn}B^{nl}=\delta _{k}^{l}, 
\end{equation}

\bigskip

where

\bigskip

\begin{equation}
B_{nm}=\left( 
\begin{array}{cc}
0 & -q^{-1/2} \\ 
q^{1/2} & 0
\end{array}
\right) ,\quad \,B^{nm}=\left(
\begin{array}{cc}
0 & q^{-1/2} \\
-q^{1/2} & 0
\end{array}
\right) ,\,  
\end{equation}

\bigskip

and $D_{q}=ad-qbc$ is the quantum determinant. $SU_{q} \left( 2 \right) $ is obtained by taking the unimodularity condition $ D_{q}=1$.

\bigskip

Let us take $q=e^{i\,\eta }\simeq 1+i\eta $. This gives

\bigskip

\begin{equation}
B_{nm}=\epsilon _{nm}+\frac{1}{2}ib_{nm},\quad B^{nm}=\epsilon ^{nm}+\frac{1%
}{2}ib^{nm}  
\end{equation}

\bigskip

where

\bigskip

\begin{eqnarray}
\epsilon _{nm} &=&\left(
\begin{array}{cc}
0 & -1 \\
1 & 0
\end{array}
\right) ,\quad \epsilon ^{nm}=\left( 
\begin{array}{cc}
0 & 1 \\ 
-1 & 0
\end{array}
\right)   \nonumber \\
b_{nm} &=&\left( 
\begin{array}{cc}
0 & \eta  \\ 
-\eta  & 0
\end{array}
\right) ,\quad b^{nm}=\left(
\begin{array}{cc}
0 & -\eta  \\
\eta  & 0
\end{array}
\right) .  
\end{eqnarray}

\bigskip

The noncommutativity of the elements $M_{\,m}^{n}$ \ is controlled by the
braiding matrix $R$

\bigskip

\begin{equation}
R=\left(
\begin{array}{cccc}
1 & 0 & 0 & 0 \\
0 & 0 & q & 0 \\ 
0 & q & 1-q^{2} & 0 \\
0 & 0 & 0 & 1
\end{array}
\right) .  
\end{equation}

$\bigskip $

$R$ becomes the permutation operator $R_{\qquad kl}^{nm}=\delta
_{l}^{n}\delta _{k}^{m}$ in the classical case $q=1.$

The $R$ matrix satisfy the Yang-Baxter equation

\bigskip

\begin{equation}
R_{\,\,\,pq}^{ij}R_{\,\,\,\,\,\,lr}^{pk}R_{\,\,\,\,\,\,mn}^{qr}=R_{\,\,%
\,pq}^{jk}R_{\,\,\,\,\,\,rm}^{ip}R_{\,\,\,\,\,\,lm}^{rq}. 
\end{equation}

\bigskip

The noncommutativity of the elements $M_{\,\,\,m}^{n}$ is expressed as

\bigskip

\begin{equation}
R_{\,\,\,\,\,\,\,nm}^{pq}M_{\,\,\,k}^{n}M_{\,\,\,l}^{m}=M_{\,\,\,n}^{p}M_{\,%
\,\,m}^{q}R_{\,\,\,\,\,\,\,\,kl}^{nm}.  
\end{equation}

\bigskip

With the nondegenerate form $B$ the $R$ matrix has the form

\bigskip

\begin{eqnarray}
R_{\qquad kl}^{+nm} &=&R_{\qquad kl}^{nm}=\delta _{\quad k}^{n}\delta
_{\quad l}^{m}+qB^{nm}B_{kl},  \nonumber \\
R_{\qquad kl}^{-nm} &=&R_{\qquad \ \ kl}^{-1nm}=\delta _{\quad k}^{n}\delta
_{\quad l}^{m}+q^{-1}B^{nm}B_{kl}.  
\end{eqnarray}

The first equation, in terms of $\eta $, gives:

\bigskip

\begin{equation}
R_{\qquad kl}^{nm}=\delta _{\quad k}^{n}\delta _{\quad l}^{m}+\epsilon
^{nm}\epsilon _{kl}+i\eta \epsilon ^{nm}\epsilon _{kl}+\frac{1}{2}\epsilon
^{nm}b_{kl}+\frac{1}{2}b^{nm}\epsilon _{kl}+o\left( \eta ^{2}\right) . 
\end{equation}

\section{Woronowicz Differential Calculus}

Now, we are going to consider the bicovariant bimodule \cite{woronowicz} $%
\Gamma $ over $SU_{q}\left( 2\right) $. Let $\theta ^{a}$ be a left
invariant basis of $_{inv}\Gamma ,$ the linear subspace of all
left-invariant elements of $\Gamma $ i.e. $\Delta _{L}\left( \theta
^{a}\right) =I\otimes \theta ^{a}.$ In the $q=1$ the left coaction $\Delta
_{L}$ coincides with the pullback for 1-forms.

There exists an adjoint representation $M_{b}^{\,\,\,\,a}$ of \ the quantum
group, defined by the right action on the left-invariant $\theta ^{a}:$

\bigskip

\begin{equation}
\Delta _{R}\left( \theta ^{a}\right) =\theta ^{b}\otimes
M_{b}^{\,\,\,\,a},\quad M_{b}^{\,\,\,\,a}\in \mathcal{A}  
\end{equation}

\bigskip

The adjoint representation is given in terms of the fundamental
representation \cite{muller} as:

\bigskip

\begin{equation}
\left( M_{b}^{\,\,\,\,a}\right) =\left( 
\begin{array}{cccc}
S\left( a\right) a & S\left( a\right) b & S\left( c\right) a & S\left(
c\right) b \\
S\left( a\right) c & S\left( a\right) d & S\left( c\right) c & S\left(
c\right) d \\
S\left( b\right) a & S\left( b\right) b & S\left( d\right) a & S\left(
d\right) b \\
S\left( b\right) c & S\left( b\right) d & S\left( d\right) c & S\left(
d\right) d
\end{array}
\right)  
\end{equation}

\bigskip

where $S \left( . \right)$ means antipode.

\bigskip

In the quantum case we have $\theta ^{a}M_{m}^{\,\,\,\,n}\neq
M_{\,m}^{\,\,\,\,n}\theta ^{a}$ in general, the bimodule structure of $%
\Gamma $ being non-trivial for $q\neq 1.$ There exist linear functionals $\
f_{\,\,\,b}^{a}:Fun\left( SU_{q}\left( 2\right) \right) \rightarrow \mathcal{%
C}$ for these left invariant basis such that

\bigskip

\begin{equation}
\theta ^{a}M_{m}^{\,\,\,\,n}=\left( \ f_{\,\,\,b}^{a}\ast
M_{m}^{\,\,\,\,n}\right) \theta ^{b}=\left( id\otimes \
f_{\,\,\,b}^{a}\right) \,\Delta \left( M_{\,m}^{\,\,\,\,n}\right) \theta
^{b}=M_{k}^{\,\,\,\,n}f_{\,\,\,b}^{a}\left( M_{\,m}^{\,\,\,\,k}\right)
\theta ^{b},  
\end{equation}

\bigskip

\begin{equation}
M_{m}^{\,\,\,\,n}\theta ^{a}=\theta ^{b}\left[ \left( f_{\,\,\,b}^{a}\circ
S^{-1}\right) \ast M_{m}^{\,\,\,\,n}\right]  
\end{equation}

\bigskip

where $ \Delta $ refers to coproduct. $ \ast $ is the convolution product of an element $ M_{m}^{\,\,\,\,n} \in \mathcal{A} $ and a functional $ \ f_{\,\,\,b}^{a} $ \cite{woronowicz}.

\bigskip 

Once we have the functionals $f_{\,\,\,b}^{a}$, we know how to commute
elements of $\mathcal{A}$ through elements of $\Gamma $. These functionals
are given by \cite{mesref}:

\bigskip

\begin{equation}
f_{\,\,\,b}^{a}\left( M_{m}^{\,\,\,\,n}\right) =q^{-\frac{1}{2}%
}R_{\,\,\,\,\,\,\,\,\,\,\,\,mb}^{an}.  
\end{equation}

\bigskip

The representation with the lower index of $\theta ^{a}$ is defined by using
the bilinear form $B$

\bigskip

\begin{equation}
\theta _{b}=\theta ^{a}B_{ab}  
\end{equation}

\bigskip

which defines the new functional $\widehat{f}_{\,\,\,b}^{a}$ corresponding
to the basis $\theta _{a}$

\bigskip

\begin{equation}
\widehat{f}_{\,\,\,d}^{c}=B_{ad}f_{\,\,\,b}^{a}B^{cb} 
\end{equation}

\bigskip

\begin{equation}
\theta _{a}M_{m}^{\,\,\,\,n}=\left( \widehat{f}_{\,\,\,a}^{b}\ast
M_{m}^{\,\,\,\,n}\right) \,\theta _{b}.  
\end{equation}

\bigskip

We can also define the conjugate basis $\theta ^{\ast a}=\left( \theta
^{a}\right) ^{\ast }\equiv \overline{\theta }_{a}$. Then the linear
functionals $\overline{f}_{\,\,\,b}^{a}$ are given by

\bigskip

\begin{equation}
\overline{\theta }_{b}M_{m}^{\,\,\,\,\,n}=\left( \ \overline{f}%
_{\,\,\,b}^{a}\ast M_{\,m}^{\,\,\,\,n}\right) \overline{\theta }_{a}
\end{equation}

\bigskip

and

\bigskip

\begin{equation}
\overline{f}_{\,\,\,b}^{a}\left( S\left( M_{\,m}^{\,\,\,\,n}\right) \right)
=q^{\frac{1}{2}}R_{\,\,\,\,\,\,\,\,\,\,\,\,mb}^{-\,an}.  
\end{equation}

\bigskip

We can easily find the transformation of the adjoint representation for the
quantum group which acts on the generators $M_{\,i}^{\,\,\,\,j}$ as the
right coaction $Ad_{R}$:

\bigskip

\begin{equation}
Ad_{R}\left( M_{i}^{\,\,\,\,j}\right) =M_{l}^{\,\,\,\,k}\otimes S\left(
M_{\,i}^{\,\,\,\,l}\right) M_{k}^{\,\,\,\,j}. 
\end{equation}

\bigskip

As usual, in order to define the bicovariant differential calculus with the $%
\ast -$structure we have required that the $\ast -$operation is a bimodule
antiautomorphism $\left( \Gamma _{Ad}\right) ^{\ast }=\Gamma _{Ad}$. We
found that the left invariant bases containing the adjoint representation
are obtained by taking the tensor product $\theta _{i}\overline{\theta }%
^{j}\equiv \theta _{i}^{\,\,\,\,j}$ of two fundamental modules. The bimodule
generated by these bases is closed under the $\ast -$operation. We found the
right coaction on the basis $\theta _{i}^{\,\,\,\,j}$

\bigskip

\begin{equation}
\Delta _{R}\left( \theta _{i}^{\,\,\,\,j}\right) =\theta
_{\,l}^{\,\,\,\,k}\otimes S\left( M_{\,i}^{\,\,\,\,l}\right)
M_{k}^{\,\,\,\,\,j}.  
\end{equation}

\bigskip

We have also introduced the basis $\theta _{ij}=\theta _{i}\,\overline{%
\theta }_{j}$

\bigskip

\begin{equation}
\theta _{ab}M_{m}^{\,\,\,\,n}=\left( \
f_{Ad\,\,\,\,\,\,\,ab}^{\,\,\,\,\,\,\,\,\,cd}\ast M_{m}^{\,\,\,\,n}\right)
\theta _{cd}=\ f_{Ad\,\,\,\,\,\,\,ab}^{\,\,\,\,\,\,\,\,\,cd}\left(
M_{m}^{\,\,\,\,k}\right) M_{\,k}^{\,\,\,\,n}\theta _{cd}  
\end{equation}

where

\bigskip

\begin{equation}
\ f_{Ad\,\,\,\,\,\,\,ab}^{\,\,\,\,\,\,\,\,\,cd}=\overline{f}%
_{\,\,\,b}^{d}\ast \widehat{f}_{\,\,\,a}^{c}.  
\end{equation}

\bigskip

The exterior derivative $d$ is defined as

\bigskip

\begin{eqnarray}
dM_{m}^{\,\,\,\,n} &=&\frac{1}{\mathcal{N}}\left[ X,M_{m}^{\,\,\,\,n}\right]
_{-}=\left( \chi ^{ab}\ast M_{m}^{\,\,\,\,n}\right) \theta _{ab}  \nonumber
\\
&=&\chi ^{ab}\left( M_{m}^{\,\,\,\,k}\right) M_{k}^{\,\,\,\,n}\theta _{ab}
\end{eqnarray}

\bigskip

where $X=B^{ab}\theta _{ab}=q^{-1/2}\theta _{12}-q^{1/2}\theta _{21}$ is the
singlet representation of $\theta ^{ab}$ and is both left and right
co-invariant, $\mathcal{N}\in \mathcal{C}$ is the normalization constant
which we take purely imaginary $N^{\ast }=-N$ and $\chi _{ab}$ are the
quantum analog of left- invariant vector fields.

Using equation (36)

\bigskip

\begin{eqnarray}
dM_{m}^{\,\,\,\,n} &=&\frac{1}{\mathcal{N}}\left( B^{ab}\theta
_{ab}M_{m}^{\,\,\,\,n}-B^{ab}\delta _{m}^{k}M_{k}^{\,\,\,\,n}\theta
_{ab}\right)  \nonumber \\
&=&\frac{1}{\mathcal{N}}\left(
B^{ab}f_{Ad\,\,\,\,\,\,\,\,ab}^{\,\,\,\,\,\,\,\,\,cd}\left(
M_{\,m}^{\,\,\,\,k}\right) M_{k}^{\,\,\,\,n}\theta _{cd}-B^{ab}\delta
_{m}^{k}M_{k}^{\,\,\,\,n}\theta _{ab}\right) .  
\end{eqnarray}

\bigskip

Then the left invariant vector field is given by

\bigskip

\begin{equation}
\chi ^{ab}=\frac{1}{\mathcal{N}}\left(
B^{cd}f_{Ad\,\,\,\,\,\,\,\,cd}^{\,\,\,\,\,\,\,\,\,ab}-B^{ab}\epsilon \right)
.  
\end{equation}

\bigskip

To construct the higher order differential calculus an exterior product,
compatible with left and right actions of the quantum group was introduced.
It can be defined by a bimodule automorphism $\Lambda $ in $\Gamma
_{Ad}\otimes \Gamma _{Ad}$ that generalizes the ordinary permutation
operator:

\bigskip

\begin{equation}
\Lambda \left( \eta _{ab}\otimes \theta _{cd}\right) =\theta _{cd}\otimes
\eta _{ab}.  
\end{equation}

\bigskip

We found \cite{mesref}

\bigskip

\begin{equation}
\Lambda _{\qquad
abcd}^{efgh}=f_{Ad\,\,\,\,\,\,\,cd}^{\,\,\,\,\,\,\,\,\,ef}\left(
M_{a}^{\,\,\,\,g}M_{b}^{\,\,\,\,h}\right)
=R_{ak}^{-fi}R_{bd}^{-kj}R_{ir}^{eg}R_{jc}^{rh}.  
\end{equation}

\bigskip

The external product is defined by

\bigskip

\begin{equation}
\theta _{ab}\,\wedge \,\theta _{cd}=\left( \delta _{a}^{e}\delta
_{b}^{f}\delta _{c}^{g}\delta _{d}^{h}-\Lambda _{\qquad abcd}^{efgh}\right)
\left( \theta _{ef}\otimes \theta _{gh}\right) .  
\end{equation}

\bigskip

The quantum commutators of \ the quantum Lie algebra generators $\chi ^{ab}$
are defined as

\bigskip

\begin{equation}
\left[ \chi ^{ab},\chi ^{cd}\right] =\left( 1-\Lambda \right) _{\qquad
efgh}^{abcd}\left( \chi ^{ef}\ast \chi ^{gh}\right) , 
\end{equation}

\bigskip

\begin{eqnarray}
\left[ \chi ^{ab},\chi ^{cd}\right] \left( M_{i}^{\,\,\,\,j}\right)
&=&\left( \chi ^{ab}\otimes \chi ^{cd}\right) Ad_{R}\left(
M_{i}^{\,\,\,\,j}\right)  \nonumber \\
&=&\chi ^{ab}\left( M_{l}^{\,\,\,\,k}\right) \otimes \chi ^{cd}\left(
S\left( M_{\,i}^{\,\,\,\,l}\right) M_{k}^{\,\,\,\,j}\right)  
\end{eqnarray}

\bigskip

where the convolution product of two functionals \cite{woronowicz} is given by:

\bigskip

\begin{equation}
\chi ^{ab}\ast \chi ^{cd}=\left( \chi ^{ab}\otimes \chi ^{cd}\right) \Delta .
\end{equation}

\bigskip

The $\chi ^{ab}$ functionals close on the quantum Lie algebra

\bigskip

\begin{equation}
\left[ \chi ^{ab},\chi ^{cd}\right] \left( M_{n}^{\,\,\,\,\,m}\right)
=\left( 1-\Lambda \right) _{efgh}^{abcd}\,\left( \chi ^{ef}\ast \chi
^{gh}\right) \left( M_{n}^{\,\,\,\,\,m}\right) ={\mathbf{C}}_{ef}^{abcd}\chi
^{ef}\left( M_{n}^{\,\,\,\,\,m}\right) , 
\end{equation}

\bigskip

where ${\mathbf{C}}_{ef}^{abcd}$ are the $q$-structure constants. They can
also be expanded in terms of $\eta $ as

\bigskip

\begin{equation}
{\mathbf{C}}_{ef}^{abcd}=C_{ef}^{abcd}+c_{ef}^{abcd},  
\end{equation}

\bigskip

where $C_{ef}^{abcd}$ is the classical matrix and $c_{ef}^{abcd}$ is the
quantum correction linear in $ \eta $.

\ The quantum Lie algebra generators $\chi ^{ab}$ satisfy the quantum Jacobi
identity

\bigskip

\begin{equation}
\left[ \chi ^{gh},\left[ \chi ^{ab},\chi ^{cd}\right] \right] =\left[ \left[
\chi ^{gh},\chi ^{ab}\right] ,\chi ^{cd}\right] -\Lambda _{\qquad
klmn}^{abcd}\left[ \left[ \chi ^{gh},\chi ^{kl}\right] ,\chi ^{mn}\right] .
\end{equation}

\bigskip

The quantum Killing metric is given by

\bigskip

\begin{equation}
g^{ab,cd}=Tr\left( \chi ^{ab}\left( M_{i}^{\,\,\,\,k}\right) \chi
^{cd}\left( M_{k}^{\,\,\,\,j}\right) \right) .  
\end{equation}

\bigskip

Let us recall that the quantum gauge theory on a quantum group $SU_{q}\left(
2\right) $ is constructed in such a way that the gauge transformations fit
the Hopf algebra structures \cite{brzezinski}. Given a left $Fun \left(
SU_{q}\left( 2\right) \right) $-comodule algebra $V$ and a quantum algebra
base $X_{B}$, a quantum vector bundle can be defined. The matter fields $%
\psi $ can be seen as sections: $X_{B}\rightarrow V$ and $V$ as fiber of $\
E\left( X_{B},\,V,\,Fun\left( SU_{q}\left( 2\right) \right) \right) $ with a
quantum structure group $Fun \left( SU_{q}\left( 2\right) \right) $.

The \ quantum Lie-algebra-valued curvature $F:Fun\left( SU_{q}\left(
2\right) \right) \rightarrow \Gamma ^{2}\left( X_{B}\right) $ is given by

\bigskip

\begin{equation}
F_{m}^{\,\,\,\,n}=F\left( M_{m}^{\,\,\,\,n}\right) =\nabla
_{m}^{\,\,\,\,\,l}\wedge \nabla
_{l}^{\,\,\,\,n}=dA_{m}^{\,\,\,\,n}+A_{m}^{\,\,\,\,\,l}\wedge
A_{l}^{\,\,\,\,n}  
\end{equation}

\bigskip

We found the infinitesimal gauge transformations \cite{mesref}

\bigskip

\begin{eqnarray}
\delta _{\alpha }A &=&-d\alpha +A\wedge \alpha =-d\alpha _{ab}\chi
^{ab}+A_{ab}\cdot \alpha _{cd}\left[ \chi ^{ab},\chi ^{cd}\right]  \nonumber
\\
&=&-d\alpha _{ab}\chi ^{ab}+A_{ab}\cdot \alpha _{cd}\left( \chi ^{ab}\otimes
\chi ^{cd}\right) Ad_{R}. 
\end{eqnarray}

\bigskip

\begin{equation}
\delta _{\beta }\left( \alpha \right) =\alpha \wedge \beta =\alpha
_{ab}\cdot \beta _{cd}\left( \chi ^{ab}\otimes \chi ^{cd}\right) Ad_{R}
\end{equation}

\bigskip

\begin{equation}
\delta _{\alpha }F=F\wedge \alpha =F_{ab}\cdot \alpha _{cd}\left( \chi
^{ab}\otimes \chi ^{cd}\right) Ad_{R}=F_{ab}\cdot \alpha _{cd}\left[ \chi
^{ab},\chi ^{cd}\right]  
\end{equation}

\bigskip

where $\alpha =\alpha _{ab}\chi ^{ab}:$ $Fun\left( SU_{q}\left( 2\right)
\right) \rightarrow X_{B}$ and where the product $\left( \cdot \right) $
denotes the exterior product of two forms on the base $X_{B}$.

We can write the last relation in terms of components as:

\bigskip

\begin{eqnarray}
\delta _{\alpha }F_{m}^{\,\,\,\,\,\,n} &=&\delta _{\alpha }F_{ab}\chi
^{ab}\left( M_{m}^{\,\,\,\,\,\,n}\right) =\left( F\otimes \alpha \right)
Ad_{R}\left( M_{m}^{\,\,\,\,\,\,n}\right)   \nonumber \\
&=&F_{cd}\chi ^{cd}\left( M_{l}^{\,\,\,\,k}\right) \otimes \alpha _{ef}\chi
^{ef}\left( S\left( M_{\,m}^{\,\,\,\,l}\right) M_{k}^{\,\,\,\,n}\right)  
\end{eqnarray}

\section{$q-$Deformed noncommutative Gauge Symmetry vs. Ordinary Gauge Symmetry}

Let us consider the quantum vector bundle $E\left( X_{B},\,V,\,Fun\left(
SU_{q}\left( 2\right) \right) \right) $ where the base space $X_{B}$ is the
Moyal plane defined through the functions of operator valued coordinates $%
\widehat{x}^{i}$ satisfying (1) and where $V$ is a left $Fun \left(
SU_{q}\left( 2\right) \right) $-comodule algebra. We define the $q$-deformed
noncommutative gauge transformations as

\bigskip

\begin{eqnarray}
\widehat{\delta }_{\widehat{\alpha }}\widehat{A} &=&-d\widehat{\alpha }+%
\widehat{A}\wedge \widehat{\alpha }=-d\widehat{\alpha }_{ab}\chi ^{ab}+%
\widehat{A}_{ab}\star \widehat{\alpha }_{cd}\left[ \chi ^{ab},\chi ^{cd}%
\right]  \nonumber \\
&=&-d\widehat{\alpha }+\widehat{A}\diamond \widehat{\alpha }-\widehat{%
\alpha }\diamond \widehat{A}  
\end{eqnarray}

\bigskip

\begin{eqnarray}
\widehat{\delta }_{\widehat{\beta }}\left( \widehat{\alpha }\right) &=&%
\widehat{\alpha }\wedge \widehat{\beta }=\widehat{\alpha }_{ab}\star
\widehat{\beta }_{cd}\left( \chi ^{ab}\otimes \chi ^{cd}\right) Ad_{R} 
\nonumber \\
&=&\widehat{\alpha }_{ab}\star \widehat{\beta }_{cd}\left[ \chi ^{ab},\chi
^{cd}\right]  \nonumber \\
&=&\widehat{\alpha }\diamond \widehat{\beta }-\widehat{\beta }\diamond 
\widehat{\alpha }  
\end{eqnarray}

\bigskip

\begin{equation}
\widehat{\delta }_{\widehat{\alpha }}\widehat{F}=\widehat{F}\wedge
\widehat{\alpha }=\widehat{F}_{ab}\star \widehat{\alpha }_{cd}\left( \chi
^{ab}\otimes \chi ^{cd}\right) Ad_{R}=\widehat{F}_{ab}\star \widehat{\alpha }%
_{cd}\left[ \chi ^{ab},\chi ^{cd}\right]  
\end{equation}

\bigskip

where $\star $ is the Groenewold-Moyal star-product defined in (2), the
convolution product $\ast $ is defined in (46) and where the new product $%
\diamond  $ is defined as

\bigskip

\begin{equation}
\widehat{\alpha }\diamond \widehat{\beta }=\widehat{\alpha }_{ab}\star 
\widehat{\beta }_{cd}\left( \chi ^{ab}\ast \chi ^{cd}\right) . 
\end{equation}

\bigskip

The classical case is obtained setting $q=1$ and $\theta =0$.

Using first order expansion in $\theta $ and $q=1+i\eta $ these relations
give

\bigskip

\begin{eqnarray}
\widehat{\delta }_{\widehat{\alpha }}\,\widehat{A}_{i\,\,ef} &=&\partial _{i}%
\widehat{\alpha }_{ef}+\left( \left[ \widehat{\alpha }_{ab}\,,\widehat{A}%
_{cd\,\,i}\right] -\frac{1}{2}\theta ^{jk}\left[ \partial _{j}\widehat{%
\alpha }_{ab}\,,\partial _{k}\widehat{A}_{icd}\right] \right) \left( C_{ef}^{abcd}+c_{ef}^{abcd}\right)   \nonumber \\
\widehat{\delta }_{\widehat{\alpha }}\,\widehat{F}_{ij\,\,ef} &=&\left( i%
\left[ \widehat{F}_{ij\,\,ab},\,\widehat{\alpha }\,_{cd}\right] -\frac{1}{2}%
\theta ^{kl}\left[ \partial _{k}\widehat{\alpha }\,_{ab},\partial _{l}%
\widehat{F}_{cd\,\,ij}\right] \right) \left( C_{ef}^{abcd}+c_{ef}^{abcd}\right) . 
\end{eqnarray}

The map between $q$-deformed noncommutative gauge field and ordinary gauge field is given by:

\bigskip

\begin{eqnarray}
\widehat{A}_{i ef}\left( A\right) &=&A_{i ef}+A_{i ef}^{\prime }\left( A\right)
=A_{i ef}-\frac{1}{4} \theta ^{k\,l} \left\{ A_{k ab},\partial
_{l}A_{i cd}+F_{li cd}\right\}  \left( C_{ef}^{abcd}+c_{ef}^{abcd}\right) +o \left( \theta ^{2}\right)  \nonumber \\
\widehat{\alpha }_{ef}\left( \alpha ,A \right) &=&\alpha _{ef} +\alpha _{ef} ^{\prime
}\left( \alpha ,A\right) =\alpha _{ef} +\frac{1}{4} \theta ^{i\,j} \left\{
\partial _{i}\alpha _{ab},A_{j cd}\right\}  \left( C_{ef}^{abcd}+c_{ef}^{abcd}\right) +o \left( \theta ^{2}\right) .
\end{eqnarray}

\bigskip 

\section{$q-$Deformed Gauge Symmetry vs. Ordinary Gauge Symmetry}

We consider now a Manin plane $x^{i}=(x,y)$, defined by $xy=qyx$, as a base space $X_{B}$ of the
quantum vector bundle. Instead of the Groenewold-Moyal star-product we use
the Gerstenhaber \cite{gerstenhaber} star product which is defined by:

\bigskip

\begin{equation}
f\star g=\mu \,\,\,e^{i\,\eta \,\,x\,\partial _{x}\otimes y\,\partial
_{y}}f\otimes g  
\end{equation}

\bigskip 

where

\bigskip 

\begin{equation}
\mu \left( f\otimes g\right) =f\,g,\qquad q=e^{i\eta }. 
\end{equation}

\bigskip 

We can write this product as:

\bigskip 

\begin{eqnarray}
f\star g &=&\sum^{\infty }_{r=0} \frac{\left( i\eta
\right) ^{r}}{r!} \left( x\frac{\partial }{\partial x}%
\right) ^{r}f\quad \left( y\frac{\partial }{\partial y}\right) ^{r}g
\nonumber \\
&=&f\,g+i\eta \,x\frac{\partial }{\partial x}f\,\,\,\,y\frac{\partial }{%
\partial y}g+o\left( \eta ^{2}\right) . 
\end{eqnarray}

\bigskip

The quantum Lie-algebra valued potential $\widehat{A}$ is given by:

\bigskip 

\begin{equation}
\widehat{A}=\widehat{A}_{ab\,i}\,dx^{i}\chi ^{ab},  
\end{equation}

\bigskip 

Using the same method we find:

\begin{equation}
\widehat{\delta }_{\widehat{\alpha }}\,\widehat{A}_{i\,\,ef}=\partial _{i}
\widehat{\alpha }_{ef}+\left( \left[ \widehat{\alpha }_{ab}\,,\widehat{A}
_{cd\,\,i}\right]   
-\eta \left( x\partial _{x}\widehat{\alpha }
_{ab}\,y\partial _{y}\widehat{A}_{icd}-x\partial _{x}\widehat{A}_{iab}
y\partial _{y}\widehat{\alpha }_{cd}\right) \right) \left( C
_{ef}^{abcd}+c_{ef}^{abcd}\right) +o\left( \eta ^{2}\right) , 
\end{equation}

\bigskip

and the new map is given by:

\begin{eqnarray}
\widehat{A}_{i ef}\left( A\right) &=&A_{i ef}+A_{i ef}^{\prime }\left( A\right) \nonumber \\
&=&A_{i ef}+ \eta \left( yA_{2 ab}xF_{1i cd} - x A_{1 ab}y \partial _{y} A_{i cd} -\partial _{i}\left( xy \right)A_{2 ab}A_{1 cd}\right) \left( C_{ef}^{abcd}+c_{ef}^{abcd} \right) +o\left(
\eta ^{2} \right),              \nonumber \\
\widehat{\alpha }_{ef}\left( \alpha ,A \right) &=&\alpha _{ef} +\alpha ^{\prime
} _{ef} \left( \alpha ,A\right) =\alpha _{ef} -\eta \left( xA_{1 ab}y\partial _{y} \alpha _{cd} \right) \left( C_{ef}^{abcd}+c_{ef}^{abcd}\right) +o \left( \eta ^{2}\right) .
\end{eqnarray}

\bigskip

We can also take a Jordanian plane as a base space of the Jordanian vector
bundle \cite{mes1,mes2}. These give a new map for the corresponding
deformed structure. 

\bigskip 

{\bf{Acknowledgements}}

I would like to thank W. R\"{u}hl and S. Waldmann for the usefull discussions and the referee for his remarks.

\bigskip

\end{document}